\documentclass[]{cta-author}

{}
{}
{}
\usepackage{units}
\usepackage{algpseudocode}
\usepackage{url}
\usepackage[utf8]{inputenc}
\usepackage{amsmath}
\usepackage[caption=false,font=footnotesize]{subfig}
\usepackage{color}

\begin{document}

\supertitle{\color{red}This paper is a preprint of a paper accepted by IET Renewable Power Generation and is subject to Institution of Engineering and Technology Copyright. When the final version is published, the copy of record will be available at the IET Digital Library.}

\title{Heuristic Optimization for Automated Distribution System Planning in Network Integration Studies}

\author{\au{Alexander Scheidler$^{1\corr}$}, \au{Leon Thurner$^{2}$}, \au{Martin Braun$^{1,2}$}}

\address{
\add{1}{Fraunhofer Institute for Energy Economics and Energy System Technology (IEE), 34119 Kassel, Germany}
\add{2}{Department of Energy Management and Power System Operation, University of Kassel, 34119 Kassel, Germany}
\email{alexander.scheidler@iee.fraunhofer.de}}

\begin{abstract}

Network integration studies try to assess the impact of future developments, such as the increase of Renewable Energy Sources or the introduction of Smart Grid Technologies, on large-scale network areas. Goals can be to support strategic alignment in the regulatory framework or to adapt the network planning principles of Distribution System Operators. This study outlines an approach for the automated distribution system planning that can calculate network reconfiguration, reinforcement and extension plans in a fully automated fashion. This allows the estimation of the expected cost in massive probabilistic simulations of large numbers of real networks and constitutes a core component of a framework for large-scale network integration studies. Exemplary case study results are presented that were performed in cooperation with different major distribution system operators. The case studies cover the estimation of expected network reinforcement costs, technical and economical assessment of smart grid technologies and structural network optimisation.
\end{abstract}

\maketitle

\section{Introduction}

\subsection{Network Integration Studies}

In recent years various system-wide network integration studies (e.g. nation-wide, state-wide, DSO-wide) have been conducted in Germany \cite{VNS, bmwi2014, vns_bayern, vns_rlp, vns_nrw, vnsh_bw17}. The goal of these studies is to estimate the expected  costs for the integration of Renewable Energy Sources (RES) in the distribution system based on different RES installation scenarios and to assess the technical and economic impact of innovative equipment and control strategies. Such system-wide network integration studies can support the strategic alignment in the regulatory framework as well as help to adapt the network planning principles of the Distribution System Operators (DSOs). A focus in these studies is:

\begin{itemize}
\item Determination of the maximum RES hosting capacity of the distribution networks in the considered area.
\item Determination of the expected costs for RES integration in the distribution system.
\item Techno-economic assessment of smart grid applications to increase the RES hosting capacity and to decrease the RES integration costs.
\end{itemize}

In the medium and low voltage level such studies are often performed on representative networks, which are chosen on the basis of structural parameters like residents per square kilometer, expected RES installation or length of feeders. The representative networks are either built as generic network models \cite{bmwi2014, vns_rlp} or representative real networks are chosen based on a cluster analysis of the network parameters \cite{VNS, vns_rlp}. A recent network integration study considers 130 network models of real HV, MV and LV networks \cite{vnsh_bw17}. The results are then extrapolated on a system-wide perspective.

Several studies at the Fraunhofer IEE and the University of Kassel that investigated a large number of real distribution networks show a considerable variation in the results and the obtained findings \cite{siw}. Especially the technical and economic assessment of different technical measures in the network planning process show a high local or regional diversity which can hardly be covered by characteristic network models. A high sample size is therefore crucial for the accuracy of the results in RES network integration studies. However, investigating a large number of networks is only possible with a high degree of automation. Data import, analysis of the research objective and export of the results all have to be automated, so that the whole process can be carried out with minimum manual engagement.

To estimate the cost of RES integration and to be able to economically compare different technologies and planning principles, it is necessary to calculate the cost of the classic network planning measures. These measures include network reconfiguration (e.g. reconfiguration of feeders), reinforcement (e.g. the increase of diameters of lines) and extension (e.g. founding new substations). Some network integration studies rely on the help of experienced network planners who plan the necessary network development for selected representative networks. However, in the scope of automation of network integration studies the calculation of the expected cost must be carried out fully automatized while it should still generate reasonable realistic solutions. The purpose of this paper is to present our automated distribution system planning framework that covers both requirements.

\subsection{Automated Distribution System Planing}

A novel property of the presented approach for automated distribution system planning is its abstract formulation that allows to apply it in different studies without great need for adaptation. Its architecture is based on three parts:
\begin{enumerate}
\item A simulation model that defines constraints and allows to evaluate violations of these constraints.
\item A number of reconfiguration, reinforcement and expansion measures that potentially solve these problems.
\item A heuristic optimization that searches for the best combination of single measures to solve the defined problems.
\end{enumerate}
The heuristic optimization does not rely on specific details of the considered measures. Instead, the heuristic is only based on the assumption that each measure can either be applied or not applied. It iteratively searches for better combinations of measures by adding and removing measures from the current best solution. While this heuristic approach does not guarantee to find globally optimal solutions, the found solutions are reasonable realistic in the scope of network studies and fulfil the purpose of estimating the expected grid reconfiguration, reinforcement and expansion costs. Additionally, the approach is much more flexible and easier to customize than e.g. analytic optimization methods. For example, it is very easy to implement and integrate new types of measures, since a mathematical formulation of the measures is not necessary. In that way, even complex measures, such as splitting a network in two parts or adding new substations, can be implemented with relatively low effort.

\subsection{Structure of the Paper}

Section \ref{grid_studies} briefly sketches the usual process of conducting automated network integration studies. Section \ref{nep} introduces the network planning problem and Section \ref{ILS} presents our approach for its solution. In Section \ref{casestudies} we present findings from four different network integration studies performed in collaboration with 4 different Distribution System Operators (DSO). Finally, in Section \ref{conclusion} we conclude and discuss possible future research questions.

\section{Automated Network Integration Studies}
\label{grid_studies}

Network integration studies usually involve one or more DSO(s), who provide the basic data that is necessary to conduct the study. Typically, the data comprises a (possibly large) number of real distribution network models. Moreover, the DSO's basic network planing rules and specific constraints are specified. This includes the definition of different worst case load cases (e.g. high feed-in case, high load case, n-1 case), the allowed network development measures and the associated cost assumptions. Additionally, different RES  scenarios, i.e. possible future development of RES in the considered area, and load development scenarios, e.g. possible future development of electric vehicles, are either provided by the DSO or developed as part of the study. Finally, the specific research questions of the study are defined, for example which Smart Grid Technologies should be investigated.

The network data is mostly provided by the DSO(s) in a commercial network calculation software format like PowerFactoy, Sincal, NEPLAN, CYME, etc. As a first step we convert all network models into the \texttt{pandapower} format. \texttt{pandapower}\footnote{www.uni-kassel.de/go/pandapower} is a Python based open source framework that is aimed at automation of power system analysis and optimization in distribution and sub-transmission networks \cite{pandapower}. The software is a joint development of the University of Kassel and the Fraunhofer IEE. It was hitherto the basis of numerous network integration studies \cite{siw, Thurner.2017}.

The possibly large number of networks in conjunction with the combinatorial complexity of the input parameters and the simulation of different probabilistic distributions of RES often leads to several thousands independent simulation runs. In order to cope with this high number mostly a HPC cluster is utilized.

A single simulation run typically investigates a specific network and a specific combination of parameters. That is, a certain random distribution of RES and loads is installed in the network model and active elements that model Smart Grid components are added. For the resulting network model it is then checked if violations of constraints in any of the considered load cases occur. In case of violations the automated network planning is started to find a combination of reconfiguration, reinforcement and extension measures that ensures safe operation of the network while being as cost-efficient as possible. The main result is an estimation of the expected network reinforcement cost for the specific network.

Finally, the results of all simulation runs are collected and statistically evaluated over the different parameter sets (RES scenarios, Smart Grid technologies, ect.).

\section{Automated Network Planning}
\label{nep}

The automated network planning is a core component of our approach to conduct large-scale network studies that involve estimating RES integration cost. In the remainder of this paper, we focus on this component. We consider a single simulation run out of possibly thousands and assume a readily parameterized network model that is in a state that violates one or more given constraints. The goal is then to find a cost-optimal set of network reconfiguration, reinforcement and expansion measures that would bring the network into a valid state. This section introduces the network planning problem and its formalization as a combinatorial optimization problem. The solution of the problem is discussed in Section~\ref{ILS}. 

In the following we introduce our approach with the help of two examples. Both examples address necessity for grid development after the installation of additional PV plants. However, the examples are only meant as an introduction in the overall framework and are by no means exhaustive. The versatility of the presented approach is showcased with the case studies presented in Section \ref{case_studies}, which, among others, also address reactive power compensation, transformer control strategies, expansion of wind power plants, and decommissioning of lines due to old age. Moreover, the assumed restrictions (e.g. allowed voltage band or the restriction to radial network structures) only serve illustration purposes and may be very different in other applications of the approach.

\subsection{An Introductory Example}
\label{example}
\begin{figure}[!htt]

\subfloat[Medium voltage line ring with 6 lines, 5 load points
and 10 load-break-switches]{	\includegraphics[width=\linewidth]{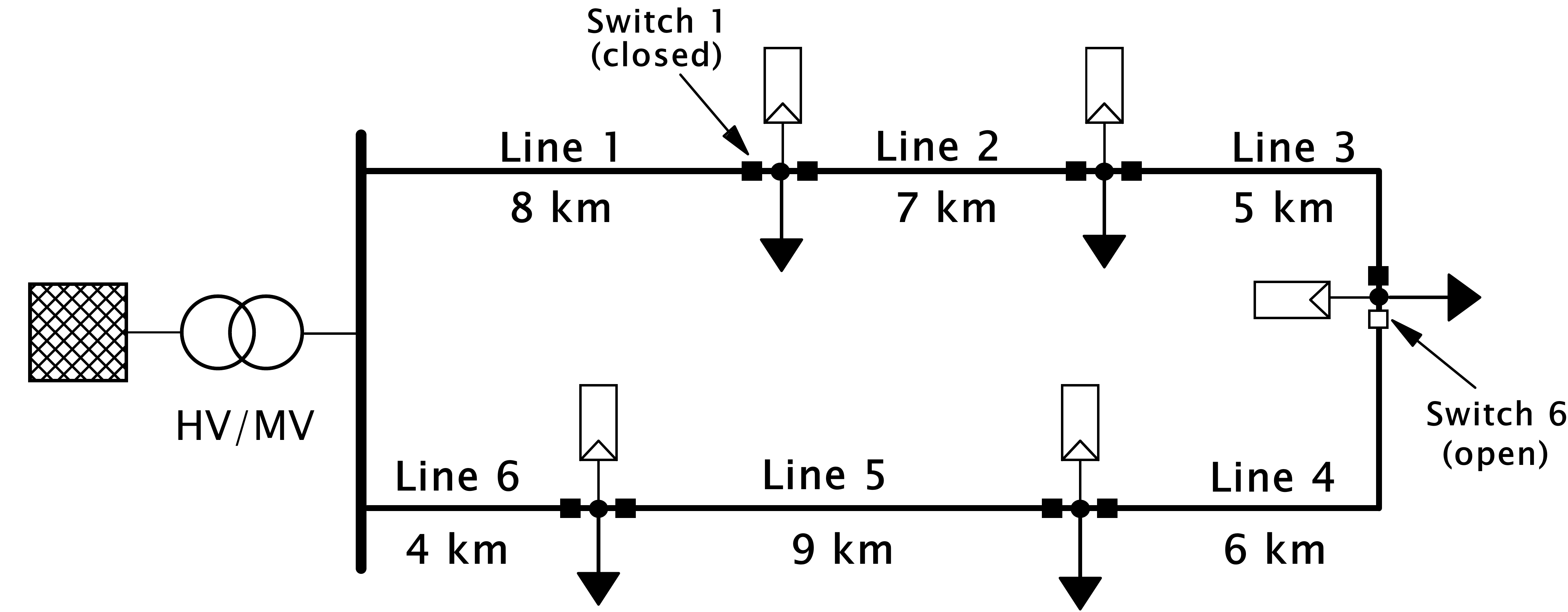}
\label{example1_net}}\\

\subfloat[Voltage profile; installation of PV plants leads to voltage band violation]{
\includegraphics[width=\linewidth]{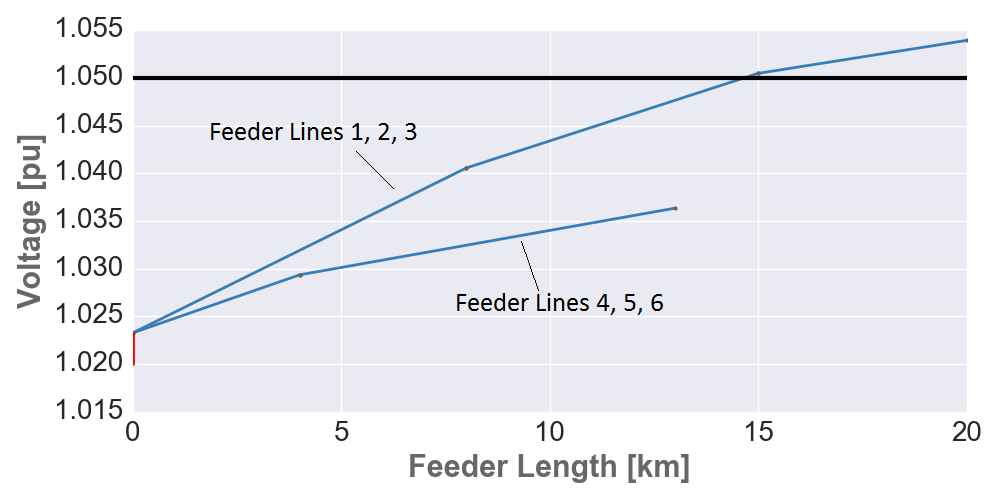} \label{example1_voltage}}

\caption{Example network}
\label{example1}
\end{figure}

The proposed approach is introduced with the help of an example. Figure \ref{example1_net} shows a medium voltage ring with 6 lines, 5 MV/LV substations modelled as load points and 10 load-break-switches. The voltage at the slack node is assumed to be 1.02 p.u. and the upper voltage limit is 1.05 p.u. Due to an increase in RES the bus voltage in the considered high feed-in scenario exceeds the voltage limit at several stations. Figure \ref{example1_voltage} shows the voltage profile of the network in this case. The vertical red line in this graph represents the voltage rise over the transformer. Note that line 4 is not depicted because the open switch is located on this line. From the network planning point of view, the network is not valid in its current state. To ensure its safe operation, the network has to be reconfigured, reinforced or extended so that all constraints are complied with.

We call every possible action that could be taken by a human network planner an abstract \emph{measure}. A measure is a single action that can be applied to the network model and that changes one or several properties of the model. A measure is atomic in the sense that it can either be applied or not. In case of the example we consider replacing existing lines by new lines with higher diameter as possible measures. The set of measures M is then given as:
\begin{equation}
M = \{\text{REPLACE\_LINE\_1, \dots, REPLACE\_LINE\_6 }\} \\
\end{equation}
We can now treat the planning problem as a combinatorial optimization problem. The overall optimization goal is to find a set of measures with minimal cost, that leads to a network state that complies with all technical constraints.

Formally, a solution $s$ is a subset of the available measures $s \subseteq M$. A solution $s$ corresponds to a network state that is the result of the application of all measures in $s$ on the initial network state. If the resulting network complies with all given constraints we call $s$ a feasible solution.

A solution has associated costs $c(s)$. In our example we define the cost of a solution as the sum of the costs $c_m$ of all individual measures in the solution:
\begin{equation}
c(s) = \sum_{m \in s} c_m
\end{equation}
Here we assume that the costs of a replacement measure is equal to the length of the replaced line.

In order to find the cost optimal network plan for the example we need to find a feasible solution $s^* \subseteq M$ that minimizes~$c(s)$: 

\begin{equation}
\begin{aligned}
& \underset{s \subseteq M}{\text{minimize}}
& & c(s) \\
& \text{subject to}
& & lp_{vv}(s) = 0,
\end{aligned}
\end{equation}

where $lp_{vv}(s)$ gives the number of load points (lp) where the voltage limit is violated after the application of $s$.

\begin{figure}[!htt]
    \centering
    \subfloat[Line 2 is reinforced with a higher diameter cable]{\includegraphics[width=\linewidth]{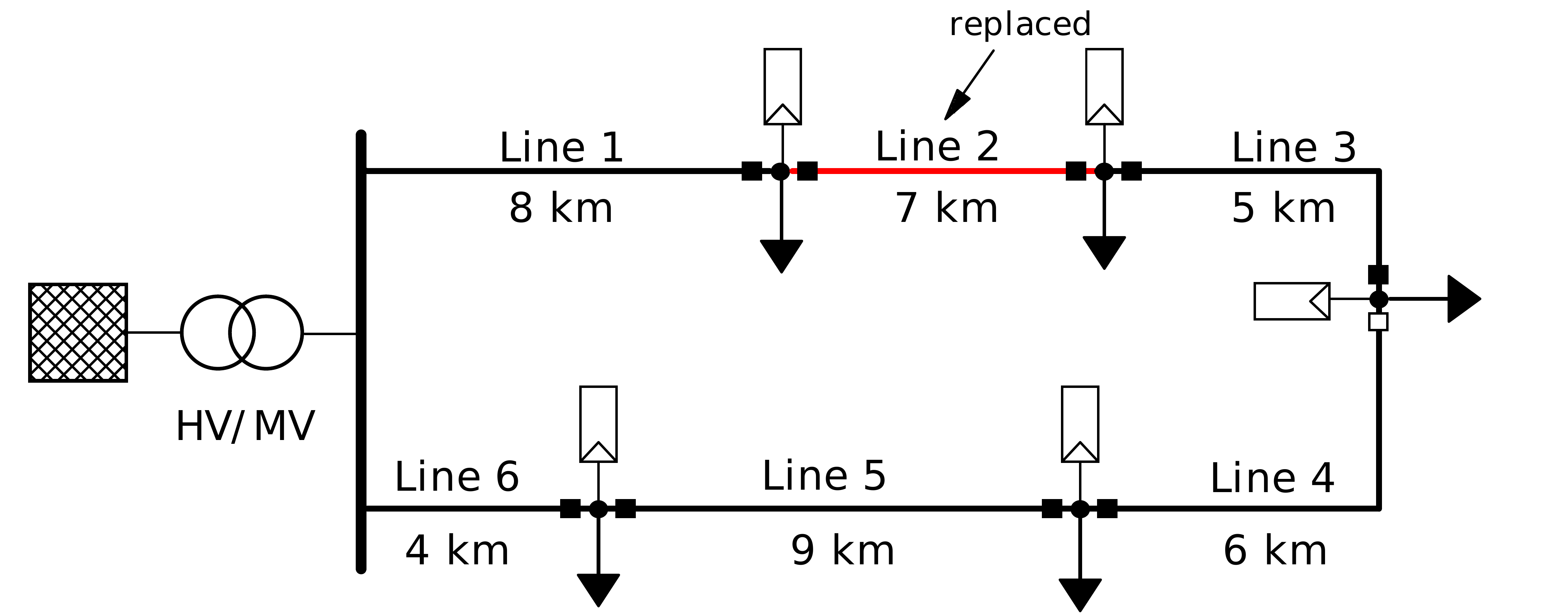}} \\
    \subfloat[Voltage Constraints are complied with after reinforcement]{\includegraphics[width=\linewidth]{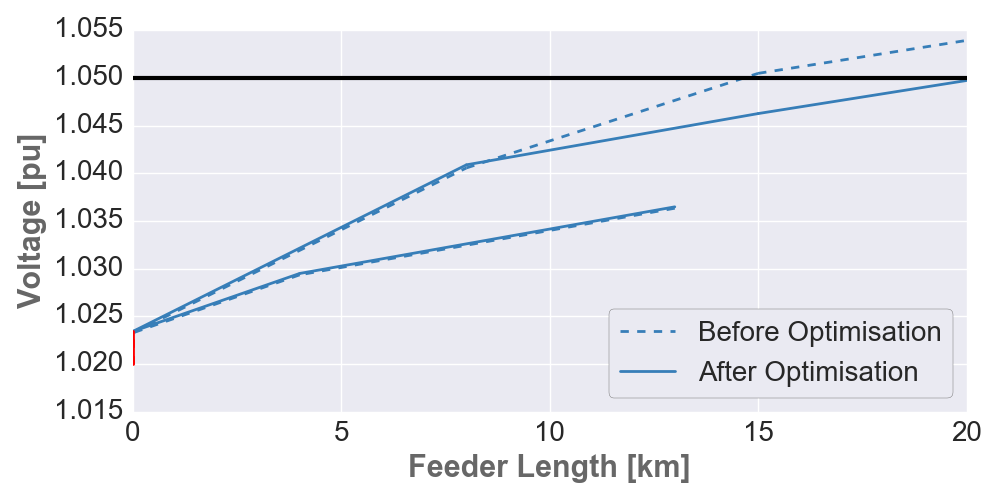}}
	\caption{Optimal reinforcement of the example network}
    \label{example_solution}
\end{figure}

The optimal solution for the example network is found by testing all $2^6$ possible combinations. This analysis yields $s^* = \{\text{REPLACE\_LINE\_2}\}$ as the most cost efficient feasible solution with cost $c(s^*) = 7$. As can be seen in Figure~\ref{example_solution}~b) all voltages are within the limits after applying $s^*$.

This solution does however not consider the possibility of moving the sectioning point to improve the voltage profile, which might lead to a more efficient solution. We therefore extend the available measures to include switching operations. We add a new degree of freedom by allowing to open a different switch than Switch 6. To this end, we simply use the network model with all switches closed as the base case and define the opening of switches as additional measures. The set of possible measures is now given as:

\begin{equation}
\begin{aligned}
M &=& \{\text{OPEN\_SWITCH\_1, \dots, OPEN\_SWITCH\_10,}\\
	&&  \text{~REPLACE\_LINE\_1, \dots, REPLACE\_LINE\_6 }\}
\end{aligned}
\end{equation}

For these 16 measures there are $2^{16}$ possible solutions. However, many of these solutions would not lead to feasible network states. For example, the solution $s = $ \{OPEN\_SWITCH\_3, OPEN\_SWITCH\_7\} would leave 2 nodes unsupplied. Opening no switch would lead to a non-radial network. To avoid this, radiality and supply constraints are considered additionally. The radiality constraint $lp_{mf}(s) = 0$ ensures that the network is not meshed (number of load points on meshed feeders equals zero). And the supply constraints $lp_{us}(s) = 0$ ensures that load points are not cut from power supply (the number of unsupplied load points equals zero). Moreover, we also add the constraint $ln_{ol}(s)=0$ to avoid overloading of lines (length of overloaded lines equals zero) and  $tr_{ol}(s)=0$ to avoid overloading of transformers (sum of transformer overloading equals zero). The optimization goal becomes:

\begin{equation}
\begin{aligned}
& \underset{s \in 2^M}{\text{minimize}}
& & c(s) \\
& \text{subject to}
& & lp_{us}(s) &= 0, \\
&&& lp_{mf}(s) &= 0, \\
&&& tr_{ol}(s) &= 0, \\
&&& ln_{ol}(s) &= 0, \\
&&& lp_{vv}(s) &= 0,
\end{aligned}
\end{equation}

\begin{figure}[!htt]
    \centering
    \subfloat[Sectioning point is moved to Switch 3 and Line 6 is reinforced]{\includegraphics[width=\linewidth]{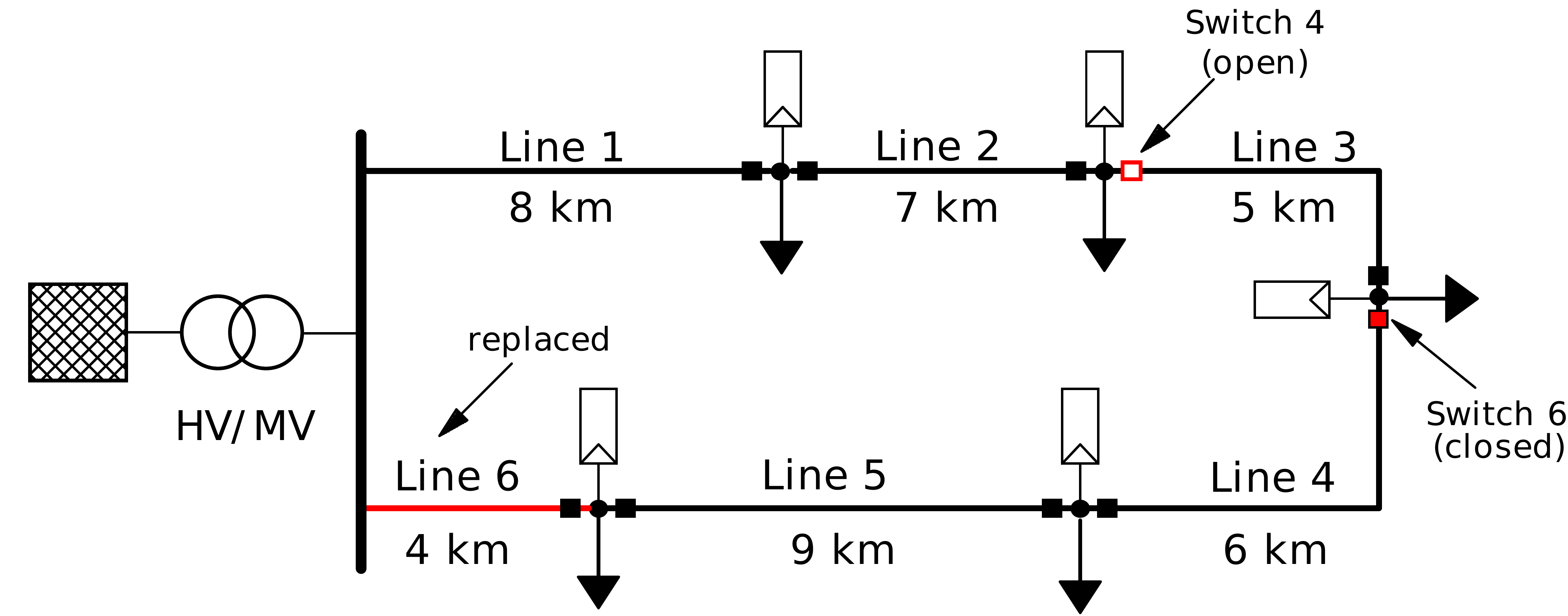}} \\
    \subfloat[Voltage Constraints are complied with after reconfiguration and reinforcement]{\includegraphics[width=\linewidth]{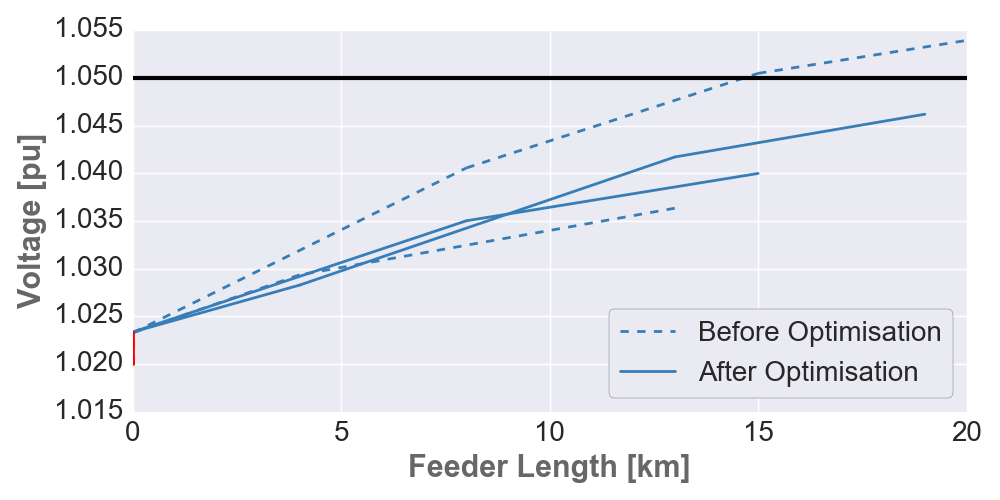}}
	\caption{Optimal reconfiguration and reinforcement of the example network}
    \label{example_solution2}
\end{figure}

The optimal solution for the example that complies with all constraints is $s^* =  \{ \text{OPEN\_SWITCH\_4, REPLACE\_LINE\_6} \}$. This solution has cost $c(s^*) = 4$ and moves the sectioning point from switch 6 to switch 4 (see Figure \ref{example_solution2}).

\subsection{A Realistic Example} \label{realistic_example}

In the introductory example the available measures were limited to exchanging lines and  opening switches. In general, the available measures are more complex and depend on the planning principles of the DSOs. In the following we present a more realistic example from one of our case studies. Figure \ref{fig_spannung} shows the example low voltage (LV) network after the installation of a large number of additional PV plants. As can be seen, several buses in the LV network are subject to violations of the voltage criteria in the considered high feed-in load case. In a first step the automated network planning component identifies the following measures that can potentially solve the voltage problems:

\begin{figure*}[!t]
\subfloat[installation of PV plants leads to voltage violation]{\includegraphics[width=0.32\textwidth]{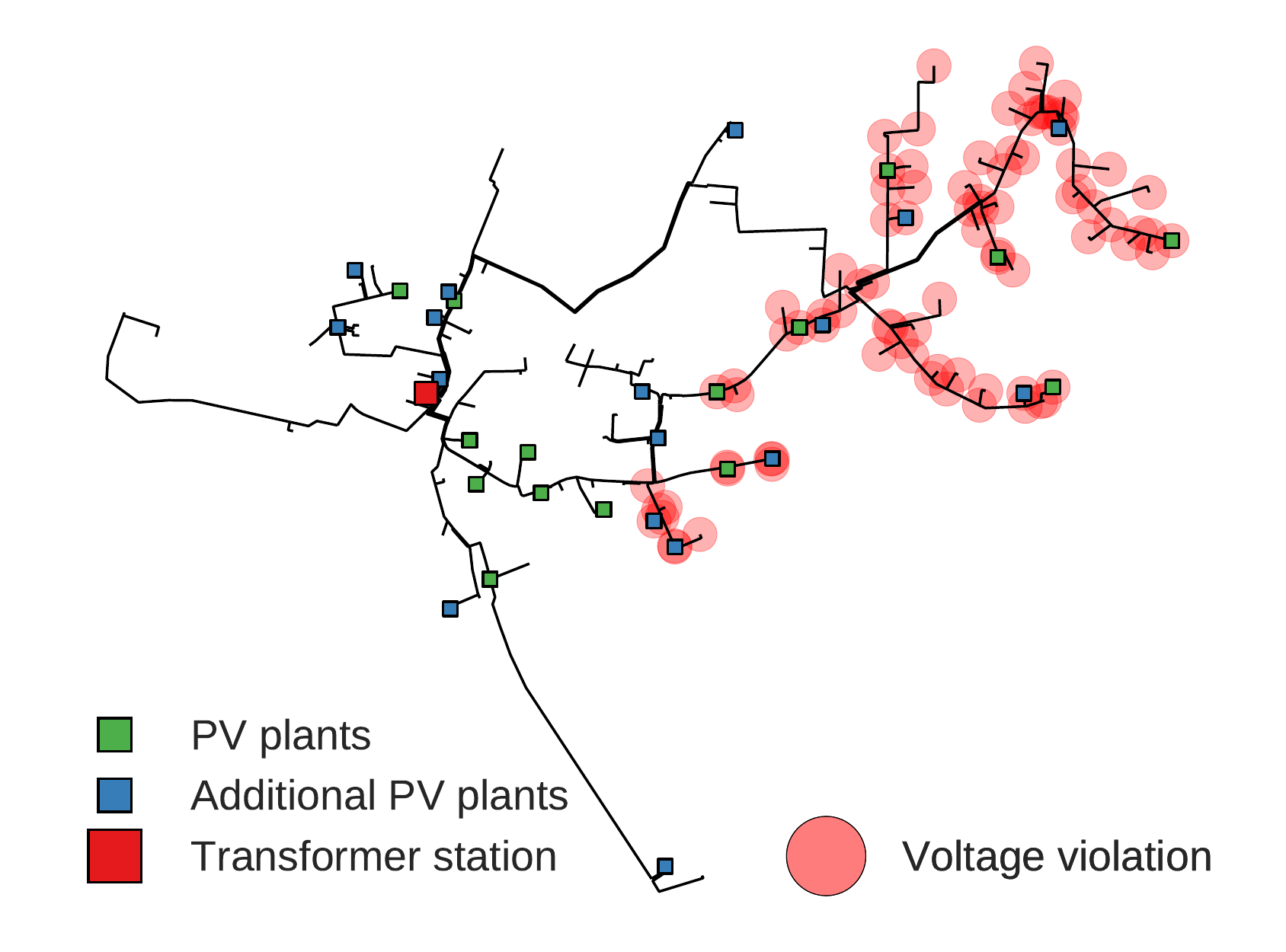}%
\label{fig_spannung}}
\hfill
\subfloat[possible line replacement measures]{\includegraphics[width=0.32\textwidth]{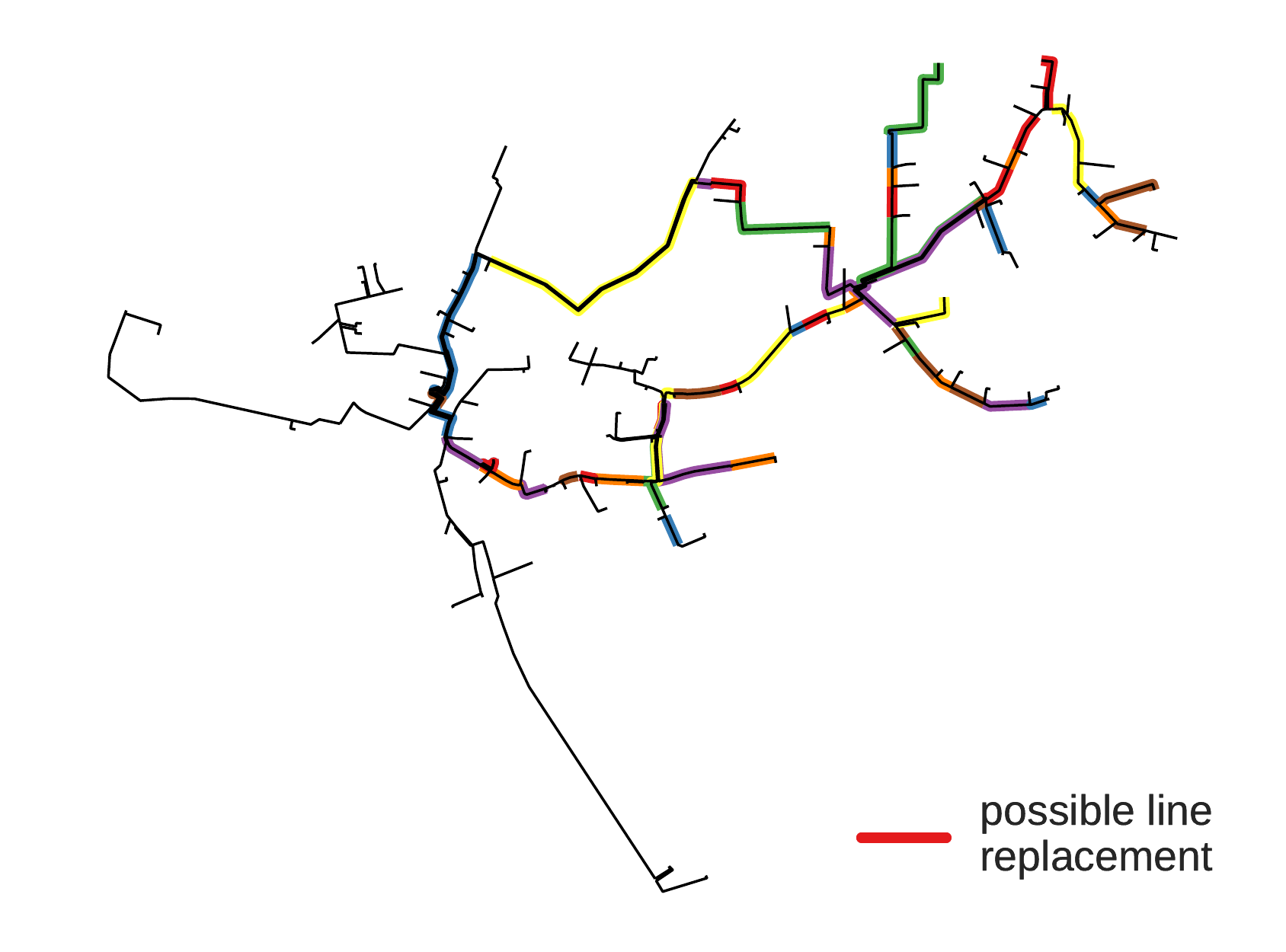}%
\label{fig_repl}}
\hfill
\subfloat[possible parallel line measures]{\includegraphics[width=0.32\textwidth]{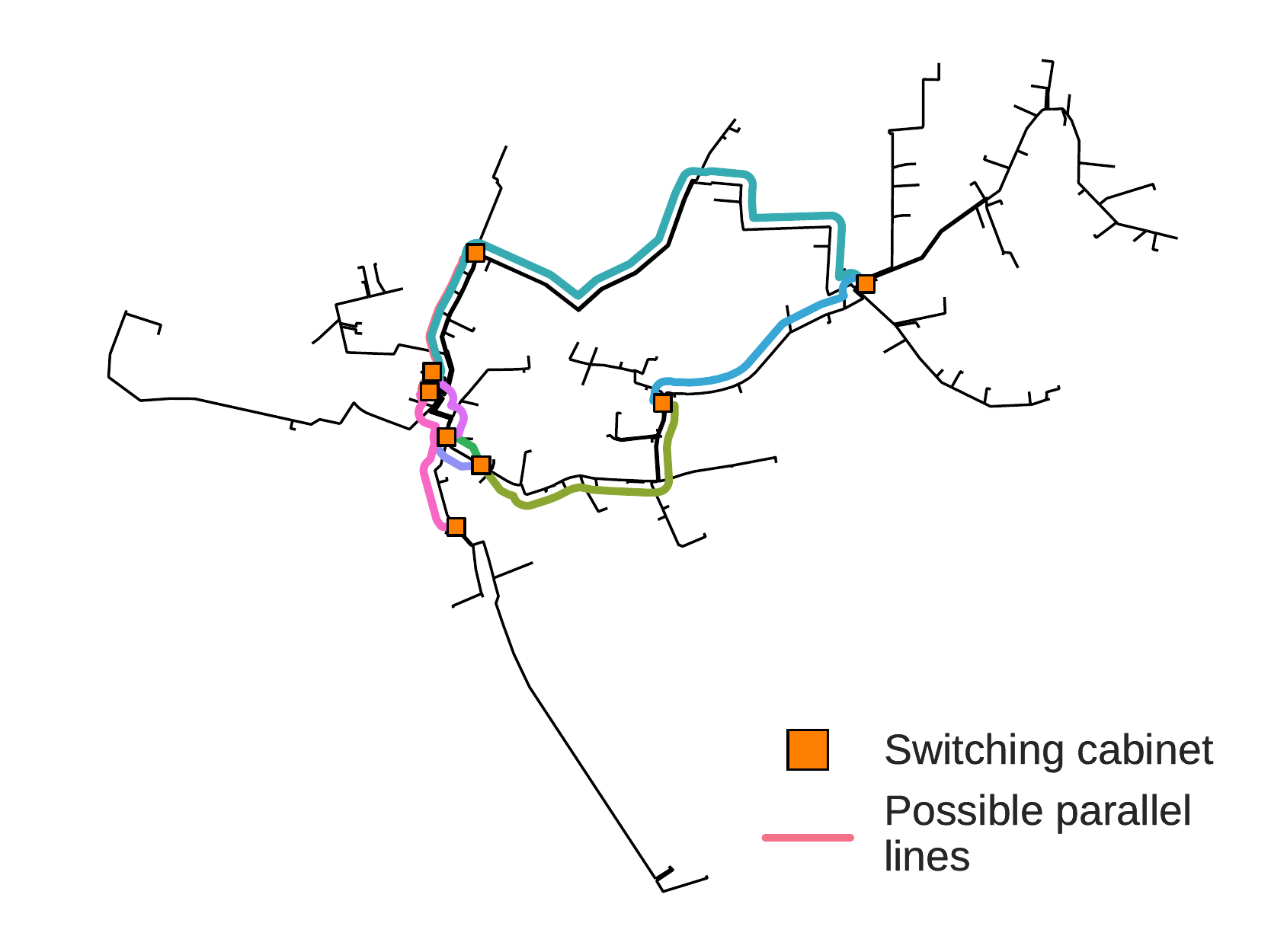}%
\label{fig_al}}\\
\subfloat[possible switch cabin + parallel line measures]{\includegraphics[width=0.32\textwidth]{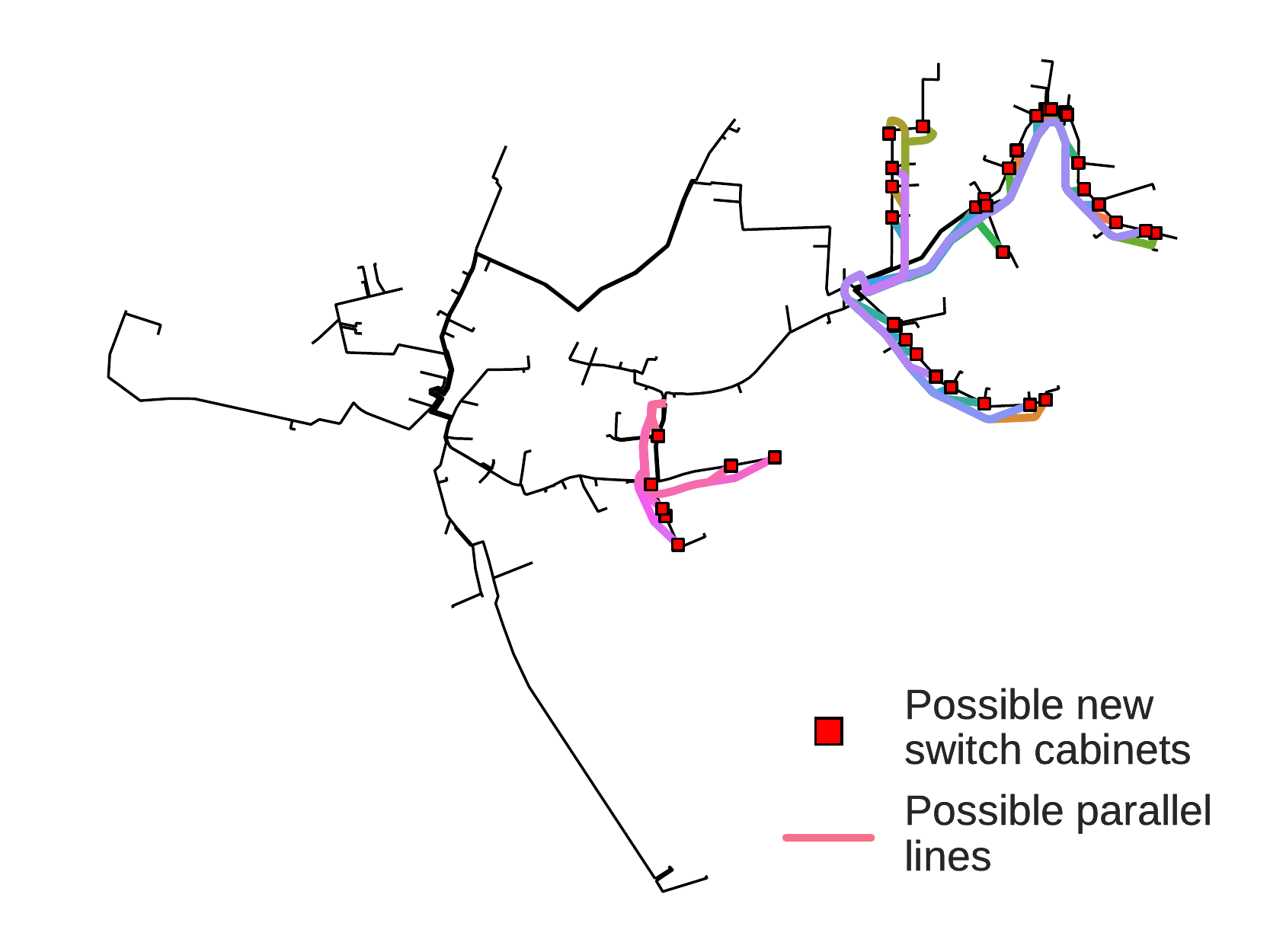}%
\label{fig_kv}}
\hfill
\subfloat[possible new station measures]{\includegraphics[width=0.32\textwidth]{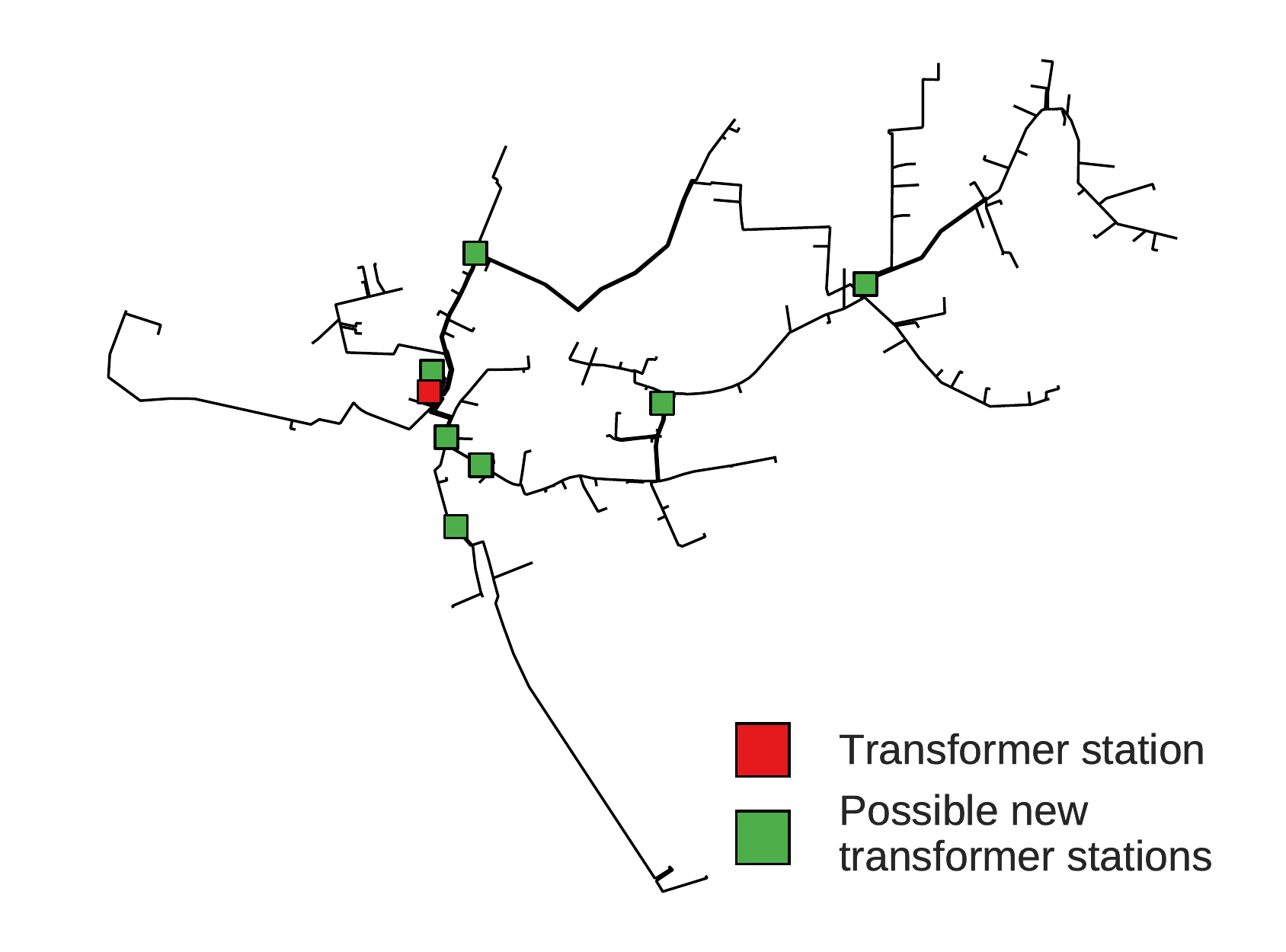}%
\label{fig_ons}}
\hfill
\subfloat[example solution]{\includegraphics[width=0.32\textwidth]{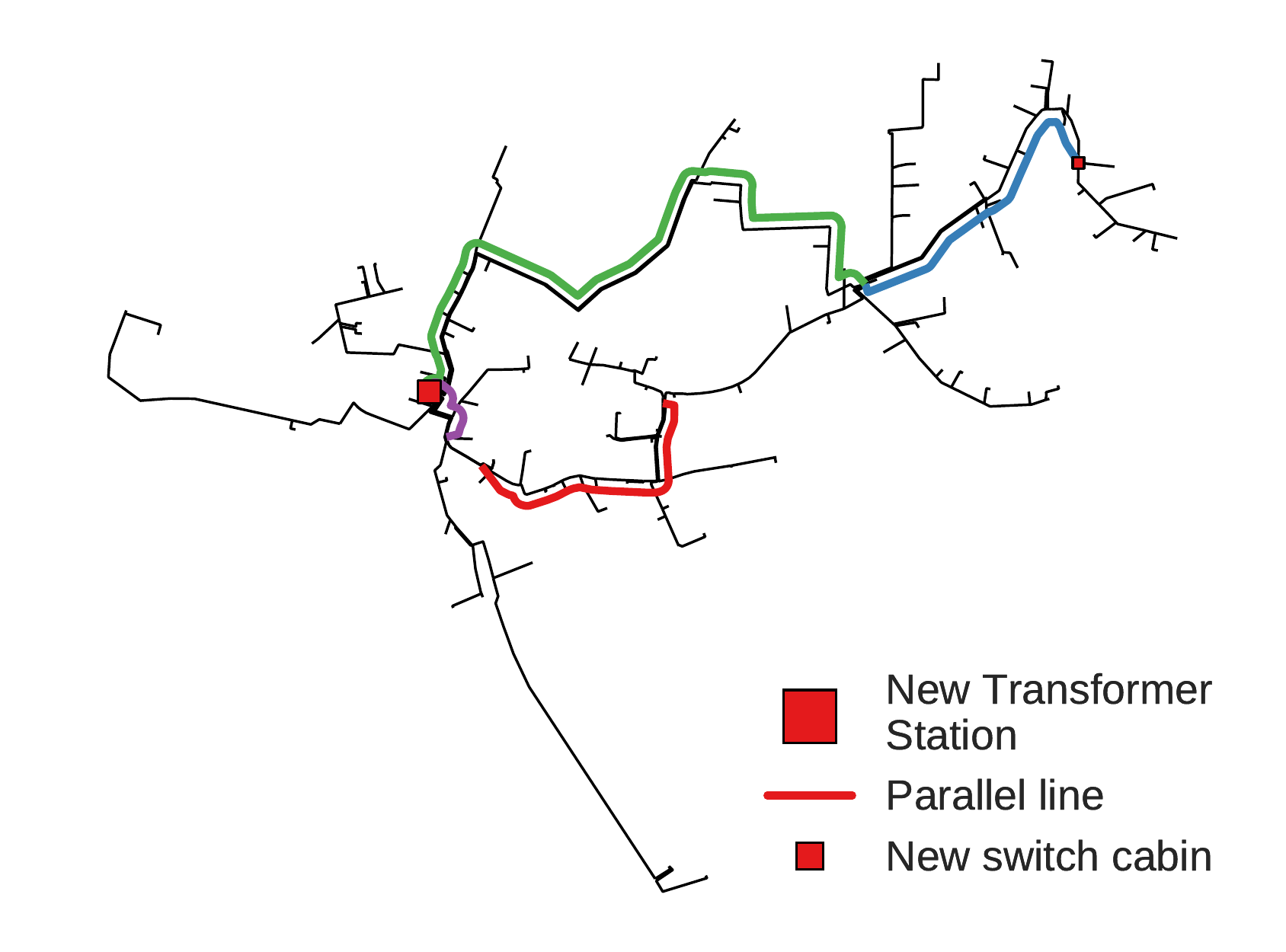}%
\label{fig_s1}}
\caption{Example Low Voltage network; depicted are different discovered measures}
\label{fig_sim}
\end{figure*}

\begin{itemize}
\item Changing the fixed tap position of the transformer. With this measure the voltage of all feeders is lowered by approximately 4 \%. However, it might lead to (more) voltage violations in the high load case.

\item Replacing cables by cables with a higher diameter. The mechanism that discovers line replacement measures only considers lines on the path from buses with voltage violations to the MV/LV transformer, since only those can effectively influence the voltage drop. Customer access lines are only considered for reinforcement if they are longer than \unit[50]{m}. The resulting options for the example network are given in Figure \ref{fig_repl}.

\item Introducing additional parallel lines between two existing switching cabinets. A requirement of the involved DSO was that parallel lines are only allowed between switching cabinets. Figure \ref{fig_al} shows the position of the switching cabinets and the parallel line measures.

\item Installing new switching cabinets plus a parallel line. In this case, the additional costs for the switching cabinet are also considered. The possible additional lines with new switch cabinets for the example network are shown in Figure \ref{fig_kv}.

\item Splitting the LV network into two networks. The position of all current switching cabinets further than \unit[50]{m} from the current substation are considered as possible locations for new substations (see Figure \ref{fig_ons}). The assumed costs include the costs for the substation itself as well as an estimate of the costs for the MV connection.

\item Replacing a transformer by a transformer of larger size. Due to the changed impedance of the transformer this measure also has influence on the voltage profile in the network.
\end{itemize}

Clearly, these kind of detailed measures can only be applied in models of real networks, since the necessary information is generally not available in generic network models. When working on realistic solutions, it is however crucial to consider all these measures and constraints in order to obtain realistic results.

Even if only measures are considered that are potentially able to mitigate the voltage problems still a total of 108 possible measures are identified. In contrast to the first example, it is not possible to find optimal solutions by brute force testing of all $2^{108}$ possible solutions. Due to the exponential growth of the solution space, an exhaustive search is generally not a feasible solution technique for realistic network planning problems.

Figure \ref{fig_s1} depicts an example solution that consists of four new parallel line measures, one new switch cabin and replacing the transformer. This solution was found using the heuristic solution approach that is presented in the next section.

\section{Heuristic Solution} \label{ILS}
The outlined network planning problem is a highly constrained, high-dimensional, mixed integer, non-linear combinatorial optimisation problem. There are numerous studies dedicated to solving network planning problems with different problem definitions and optimisation approaches \cite{Georgilakis201589, Jordehi20151088, Sedghi2016415, Zubo20171177, Ganguly2013}. One approach is to use deterministic algorithms such as linear programming, non-linear programming and dynamic programming \cite{Franco2014265}. While deterministic solvers guarantee to find the global optimum, they usually require simplifications in the constraints, reduction of the solution space or linearisation of the problem to create a problem that can be solved analytically. The realistic example presented in Section \ref{realistic_example} demonstrates that a realistic network planning problem is too complex and non-linear for analytical approaches. Metaheuristics, on the other hand, neither require differentiability, continuity, nor convexity of objective functions and are efficient in handling constrained, discrete and multi-modal optimisation problems. They are therefore very popular in solving distribution system planning problems. Solving the grid planning problem with
a metaheuristic approach was pioneered by V. Miranda in 1994 \cite{Miranda1994}. Since then, Different metaheuristics were successfully applied to solve network planning problems, such as genetic algorithms~\cite{Miranda1994, Chen.2000, Camargo2013, Fletcher.2015, Falaghi2011, Kong.2009}, particle swarm optimisation \cite{Sahoo2012, Ganguly2010}, tabu search \cite{Cossi2010, Navarro.2009.2}, artificial immune systems \cite{Keko2004, Lin.2007}, Iterated Local Search (ILS) \cite{siw} and evolutionary algorithms \cite{Zmijarevic2005, Cossi.2005, Gitizadeh.2013}. Heuristic algorithms have also been successfully applied to line routing problems that take the grid topology and geography into account \cite{Domingo.2011,Keko2004,Zmijarevic2005,Kong.2009, Navarro.2009.2}.

In this paper, we focus on so-called stochastic local search metaheuristics. These algorithms follow a single chain of solutions through the solution space searching for good solutions \cite{Lourenco2003}. They usually consist of three parts: a method to determine initial solutions $s_0$, a neighbourhood function $N(s)$ that defines how new candidate solutions are generated from the actual solution $s$, and an acceptance criteria that is used to decide if a (randomly) selected neighbour $s' \in N(s)$ is accepted as the new actual solution.

As initial solution $s_0$ we usually choose the empty set that represents the initial network. Neighbourhood function, cost function and different algorithms are outlined in the following.

\subsection{Neighbourhood Function}
Different variants exist for the neighbourhood function. We mostly define $N(s)$ as all solutions that can be derived from $s$ by either removing a single measure $m \in s$, adding a single measure $m \not \in s$ or exchanging a measure $m_1 \in s$ with a measure $m_2 \not \in s$. However, depending on the heuristic and problem it can also be beneficial to only allow adding and removing but not exchanging measures.

The neighbourhood can be further restricted by defining dependencies between measures. This allows to enforce additional constraints or to reduce the size of the neighbourhood. Possible dependencies are:
\begin{itemize}
\item a measure can exclude other measures (e.g., opening only one switch on the same line section)
\item a measure can require other measures (e.g., a parallel line to a given line is only allowed if this line is already replaced by a line with maximal diameter)
\item at least one of a set of measures must be included in the solution (e.g., a line has to be exchanged because of its age)
\end{itemize}

\subsection{Cost Function}
Most solutions in the solution space are not feasible since they violate one or more constraints. One method to cope with non-feasible solutions would be to restrict the search process to feasible regions in the solution space. However, this implies several difficulties. First of all, one would require means to generate feasible starting solutions. Yet, this is often a hard task in itself. Moreover, regions of feasible solutions in the solution space do not need to be connected with respect to the neighbourhood function and consequently the search might never reach certain regions in the solution space.

Instead of excluding non-feasible solutions, we extend the cost function with the intend to guide the search process towards feasible solutions. To this end, the degree of constraint violation is included in the cost function. More precisely, the cost function is defined as a tuple of two values, where the first value of this tuple is a number that represents the violated constraint and the second value represents the strength of the violation. In case no constraint is violated the first value is zero and the second gives the cost $c(s)$ of the solution. The cost of two solutions $c'(s_1) = (p_1, v_1)$ and $c'(s_2) = (p_2, v_2)$ is then compared lexicographically, that is, $(p_1, v_1) < (p_2, v_2)$ iff $(p_1 < p_2) \vee (p_1 = p_2 \wedge v_1 < v_2$).

The extended cost function for the example would then be:

\begin{equation}
  c'(s)=\left\{
  \begin{array}{@{}ll@{}}
    (5, lp_{us}(s)), & \text{if  } lp_{us}(s) > 0 \\
    (4, lp_{mf}(s)), & \text{if  } lp_{mf}(s) > 0 \\
    (3, tr_{ol}(s)), & \text{if  } tr_{ol}(s) > 0 \\
    (2, ln_{ol}(s)), & \text{if  } ln_{ol}(s) > 0 \\
    (1, lp_{vv}(s)),  & \text{if  } lp_{vv}(s) > 0 \\
    (0, c(s)), & \text{otherwise}
  \end{array}\right.
\end{equation}

\subsection{Stochastic Local Search}
Different stochastic local search algorithms can be implemented based on the same cost and neighbourhood functions. 

\subsubsection{Hill Climbing (HC)}

One of the simplest stochastic local search algorithms is Hill Climbing. Hill Climbing starts with an initial solution $s_0$ and iteratively moves to a random neighbouring solution if this step decreases the cost function. This is repeated until no improving step is available any more. As pseudo-code, the algorithm can be written as:

\vspace{0.3em}
\begin{algorithmic}[1]
\Procedure{HillClimbing}{$s_0$}
\State $s^{*} \leftarrow s_0$
\Repeat
\State Choose $s' \in N(s^*)$
\If{$c(s') < c(s^*)$}
\State $s^* = s'$
\EndIf 
\Until{$c(s) \geq c(s^*), \forall s \in N(s^*)$}
\State \textbf{return} $s^*$
\EndProcedure
\end{algorithmic}
\vspace{0.3em}

\subsubsection{Iterated Local Search}

Clearly, Hill Climbing will quickly end up in a local optimum. A valid approach for finding better solutions is then to just restart Hill Climbing several times. However, this would "forget" all information collected in the search process so far. A different approach is to have means for a limited acceptance of worsening moves. For example, to escape local optima by slightly perturbing the actual solution. Iterated Local Search (ILS) does exactly this. In general, the search for better solutions in ILS occurs in the reduced solution space defined by an arbitrary black-box heuristic. That is, ILS moves through the solution space from a local optimum to neighbouring (improved) local optima. We use Hill Climbing algorithm as local search algorithm for the ILS.

\vspace{0.3em}
\begin{algorithmic}[1]
\Procedure{IteratedLocalSearch}{$s_0$} 
\State $s^{*} \leftarrow HillClimbing(s_0)$
\While{not stopping criteria met} 
\State $s' \leftarrow Perturbate(s^*)$
\State $s^{*'} \leftarrow HillClimbing(s')$
\If{$c(s^{*'}) < c(s^*)$}
\State $s^* = s^{*'}$
\EndIf 
\EndWhile
\State \textbf{return} $s^*$
\EndProcedure
\end{algorithmic}
\vspace{0.3em}
To perturbate a solution we simply move $strength$ of arbitrary steps in the neighbourhood of the solution:
\vspace{0.3em}
\begin{algorithmic}[1]
\Procedure{perturbate}{$s'$, strength}
\For{$strength$ steps}
\State $s' \leftarrow \text{Choose solution} \in N(s')$
\EndFor
\State \textbf{return} $s'$
\EndProcedure
\end{algorithmic}\vspace{0.3em}

For the stopping criteria different variants can be used. In network studies we usually stop ILS after a fixed number of iteration without improvement.

\subsubsection{Late Acceptance Hill Climbing}

Late Acceptance Hill-Climbing (LAHC) is a local search algorithm, which accepts non-improving moves when a candidate cost function is better than it was a number of iterations before \cite{Burke201770}. It has been shown that LAHC performs comparably to other well established heuristics like Simulated Annealing (SA) \cite{Kirkpatrick83optimizationby}. While SA and similar techniques follow a cooling schedule that includes to calculate the cost difference in case of worsening moves, this is not necessary in LAHC. This fact plays well with our definition of the cost function (it is not meaningful to quantify the difference between the cost values of two solutions that lead to different constraint violations).

\begin{figure}[!htt]
\subfloat[Evolution of cost function values of the actual solution for different Metaheuristics]{\includegraphics[width=\linewidth]{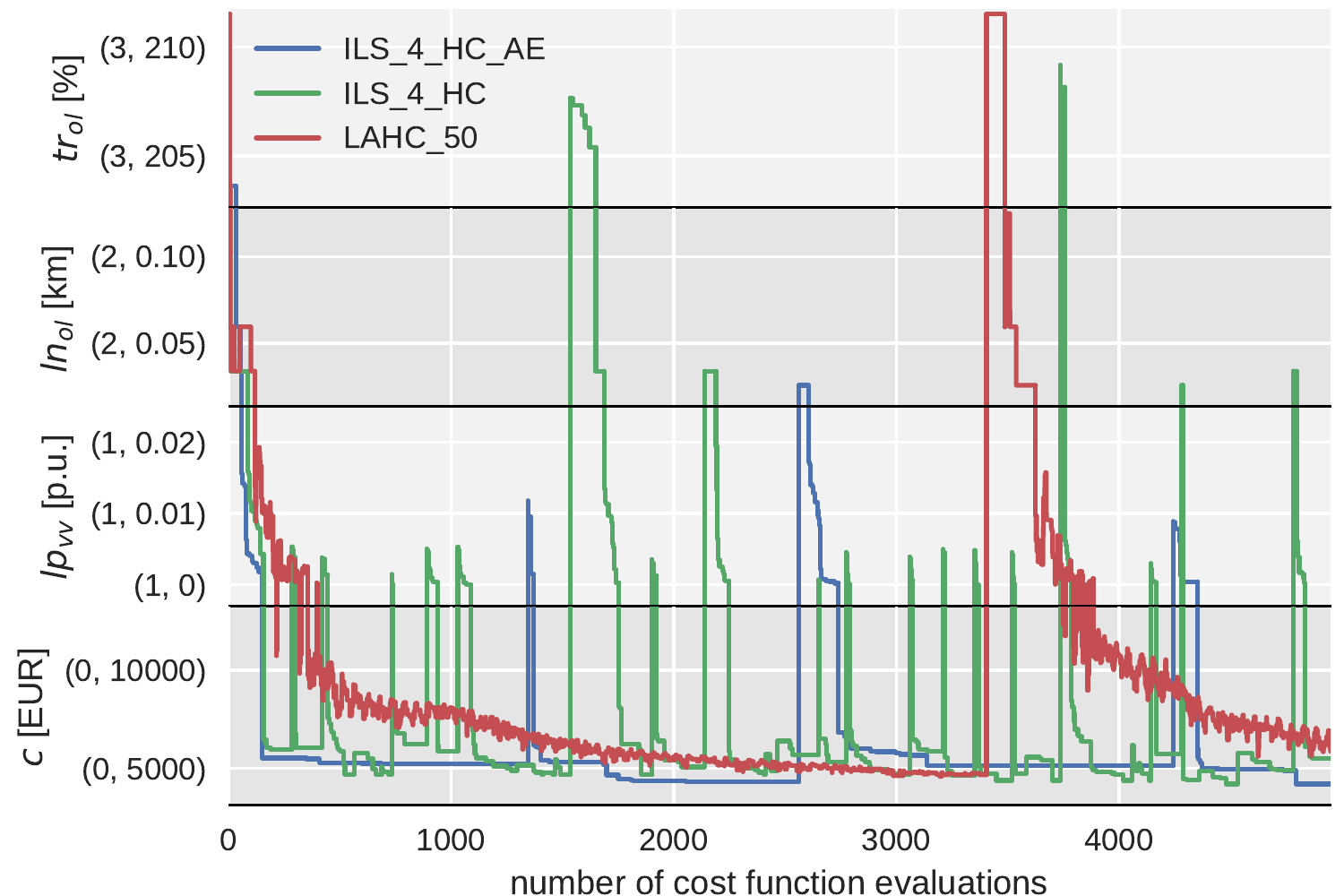}%
\label{fig_fitness}}
\\
\subfloat[Comparison of the distributions of the best found solutions (50 runs, 5000 fitness evaluations) for two different networks]{\includegraphics[width=\linewidth]{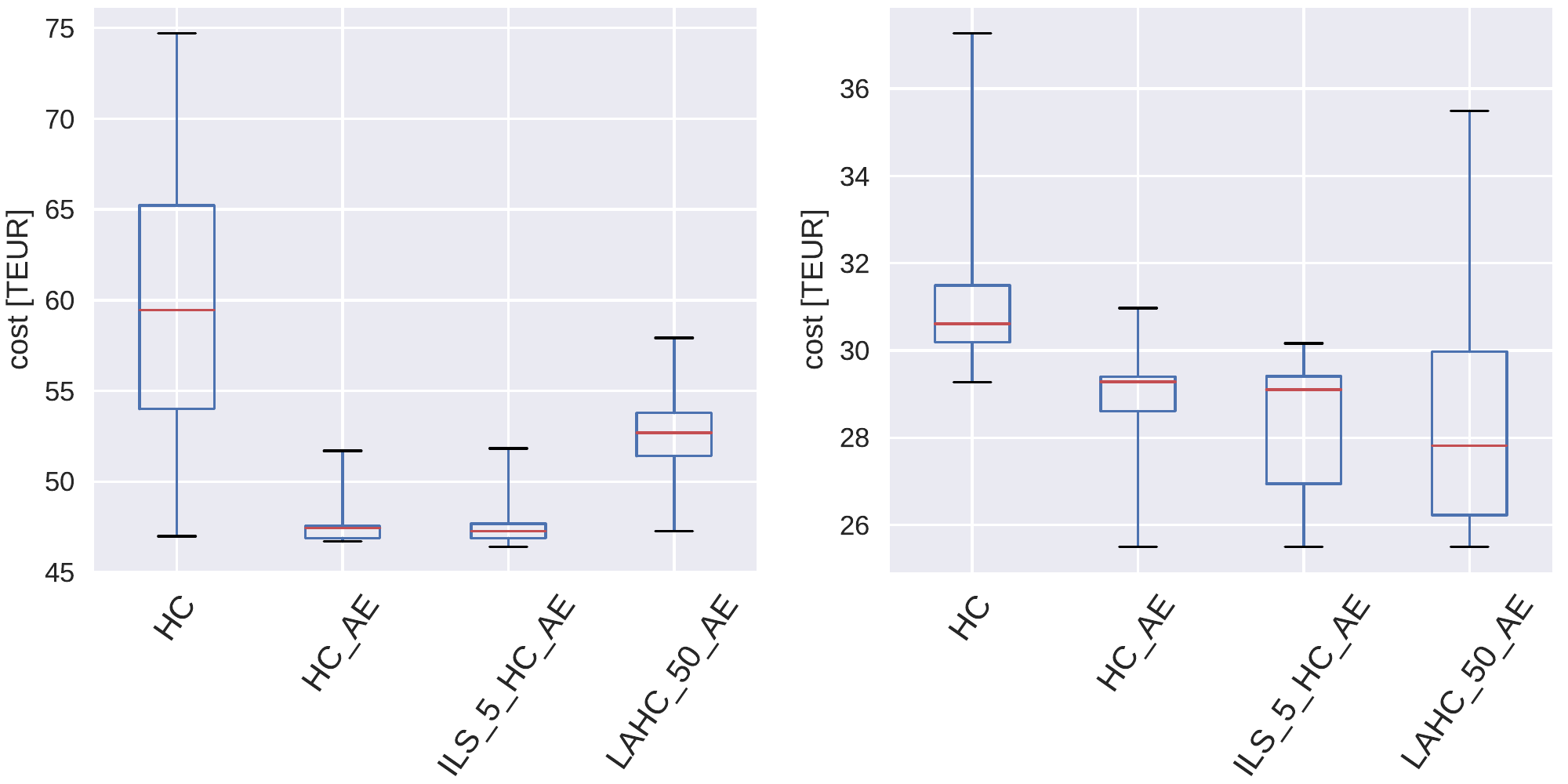}%
\label{fig_heur_comp}}
\caption{Comparison of different Metaheuristics}
\label{fig_heur}
\end{figure}

\subsection{Comparison of Algorithms}
Figure \ref{fig_fitness} depicts how the extended cost function of the actual solution evolves over time for different metaheuristics. One of our real network case studies is used to compare the following three algorithms:
\begin{itemize}
\item ILS\_4\_HC: ILS with HC as local search, perturbation is done with 4 random steps in the neighbourhood. The neighbourhood consist of adding and removing measures.
\item ILS\_4\_HC\_AE: same as ILS\_4\_HC, only that the neighbourhood additionally allows exchange (AE) of measures.
\item LAHC\_50: LAHC with a threshold of 50 previous solutions 
\end{itemize}
The y-axis is divided into 4 areas that represent the 4 different values at the first position in the cost tuple: transformer overloading, line overloading, voltage violations and cost. For the ILS algorithms, perturbations are carried out repeatedly until 5000 cost function evaluations are reached. Between two ILS iterations, the fitness decreases monotonic due to the Hill Climbing local search. The perturbating step on the other hand usually leads to a worsening in the cost function. In the depicted 5000 fitness evaluations only 3 perturbation steps occur for ILS\_4\_HC\_AE whereas for ILS\_4\_HC more than 20 can be observed. This is due to the fact that the neighbourhood of ILS\_4\_HC\_AE is significantly larger than in ILS\_4\_HC due to the allowed exchange of measures. The LAHC heuristic always allows worsening steps and therefore takes more than 3000 cost function evaluations for LAHC to reach a local optimum. It is restarted until 5000 fitness evaluations are reached.

Which metaheuristic is best suited to solve a problem can only be decided based on the specific use case. Figure \ref{fig_heur_comp} shows results from two real networks, using the same type of considered measures and cost function. The only difference is the network, which also has an effect on the number of measures and violated constraints. It can be seen that the different heuristics perform differently in each network. While in one network a restarted Hill Climbing with allowed measure exchange performs on par with the ILS variant, for the other network the LAHC variant is the best. This shows that there is not one metaheuristic that clearly outperforms all other algorithms. It is however possible to draw conclusions from comparing different algorithms. For example, it seems like the AE algorithms generally outperform the non AE algorithm. At least for this problem, allowing an exchange of measures seems to pay off, even though it significantly increases the neighbourhood. Larger sample sizes with more heterogeneous problems are needed to deduce more general conclusions about different algorithms.

\section{Case Studies} \label{case_studies}
\label{casestudies}
The automated network planning framework allows flexible combination of measures, constraints and cost functions to reflect different objects of investigation. Within different studies we developed measure implementations for, e.g.:
\begin{itemize}
\item replacing existing lines and transformers
\item adding parallel lines to existing line trails
\item changing the switching configuration
\item finding new line trails
\item deploying advanced control functions to transformers and PV systems
\item replacing conventional transformers with On-Load-Tap-Changing (OLTC) transformers
\end{itemize}

Moreover, in different studies we considered the following constraints:
\begin{itemize}
\item radiality, supply, n-1 and other topological constraints
\item load flow constraints for bus voltage, line loading, transformer loading
\item several worst-case scenarios, e.g. high load or high generation, simultaneously
\item load flow constraints for n-1 operation with optimal resupply
\item reliability constraints for outage times (ASIDI / SAIDI)
\end{itemize}

In the following we present exemplary results from four network studies that were performed in cooperation with different major distribution system operators. Due to lack of space many details have to be neglected. However, we think the case studies allow a general understanding of what kind questions can be answered with the presented approach. The metaheuristics that are used in the studies are adapted and tuned versions of the basic variants presented in Section \ref{ILS}. Specific cost values are omitted for confidentiality reasons.

\subsection{Expected Network Reinforcement Cost for LV Networks}
One goal of this study was to estimate the expected network reinforcement cost for different RES scenarios. The study was done in cooperation with the German DSO EnergieNetz Mitte.

\begin{figure}[!htt]
    \centering
    \includegraphics[width=\linewidth]{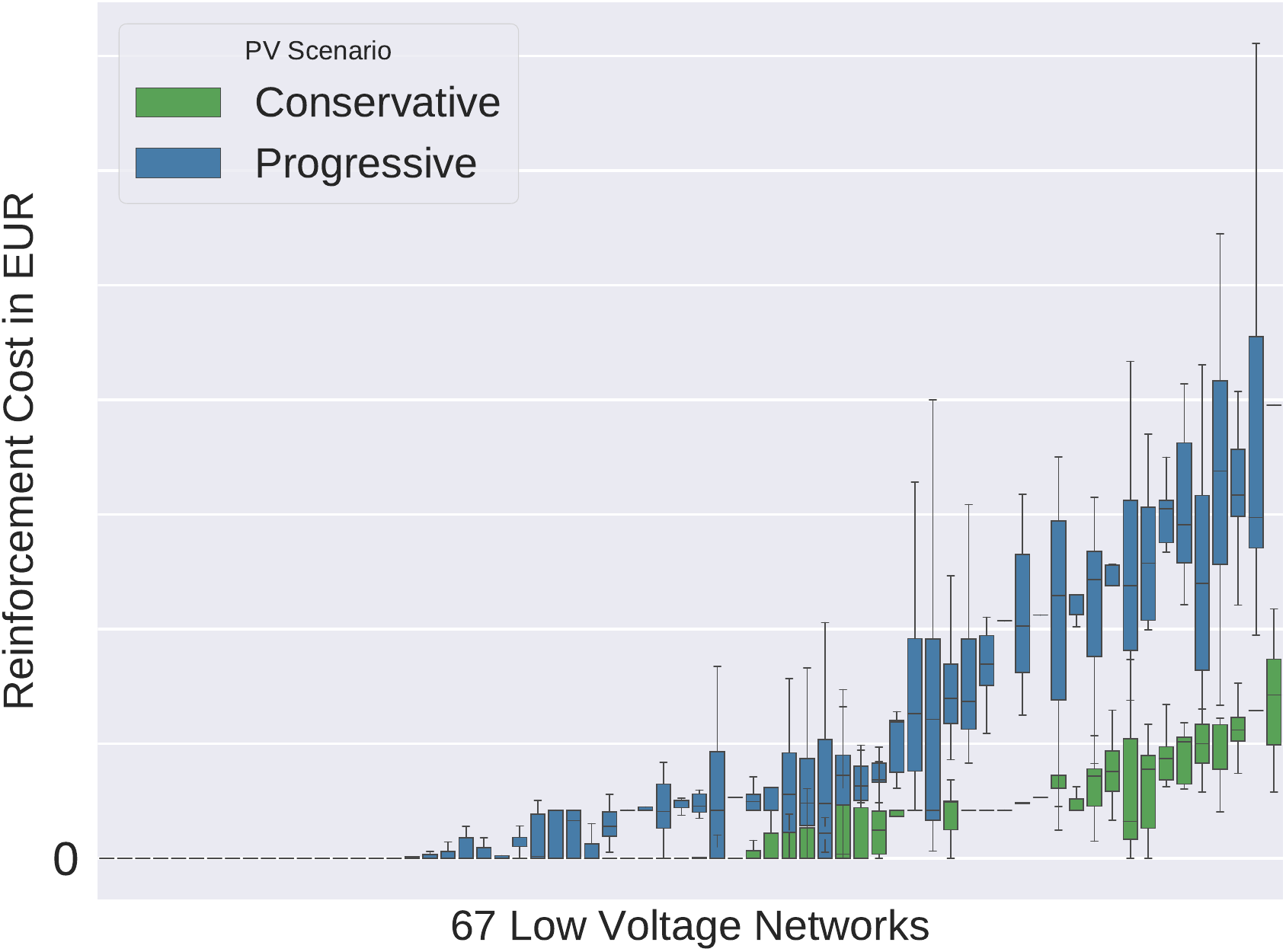}
	\caption{Example results of the expected reinforcement cost for 2 different RES scenarios}
    \label{fig_vnsh}
\end{figure}

Figure \ref{fig_vnsh} shows results for two different PV scenarios. The distribution of the cost values for the single LV networks is the result of the simulation of 50 different probabilistic distributions of PV plants within each network. As can be seen the different networks have different sensitivities to the specific distributions of PV plants. The results show moreover that for the conservative scenario half of the investigated 67 LV networks (with a total of 570 feeders) need reinforcement measures. In the progressive scenario more than two thirds of the networks are affected and the expected costs per network are higher.

\subsection{Inverter Control Strategies in MV Networks}

Within the scope of a PV integration study carried out for the Swiss distribution system operator Romande Energie, the hosting capacity of 111 LV networks and two MV networks was evaluated and compared to different PV forecast scenarios provided by the DSO (for a definition of the term hosting capacity see e.g. \cite{siw}). The expected PV capacity in 2035 exceeds the determined PV hosting capacity of the LV networks only in few cases. Therefore, the LV networks are mostly well dimensioned for the expected additional PV installations and additional measures for PV integration are just expected in a few LV networks.

\begin{figure}[!htt]
\subfloat[Comparison of the expected network reinforcement cost using different Smart Grid technologies for a MV network of the swiss DSO Romande Energy]{\includegraphics[width=\linewidth]{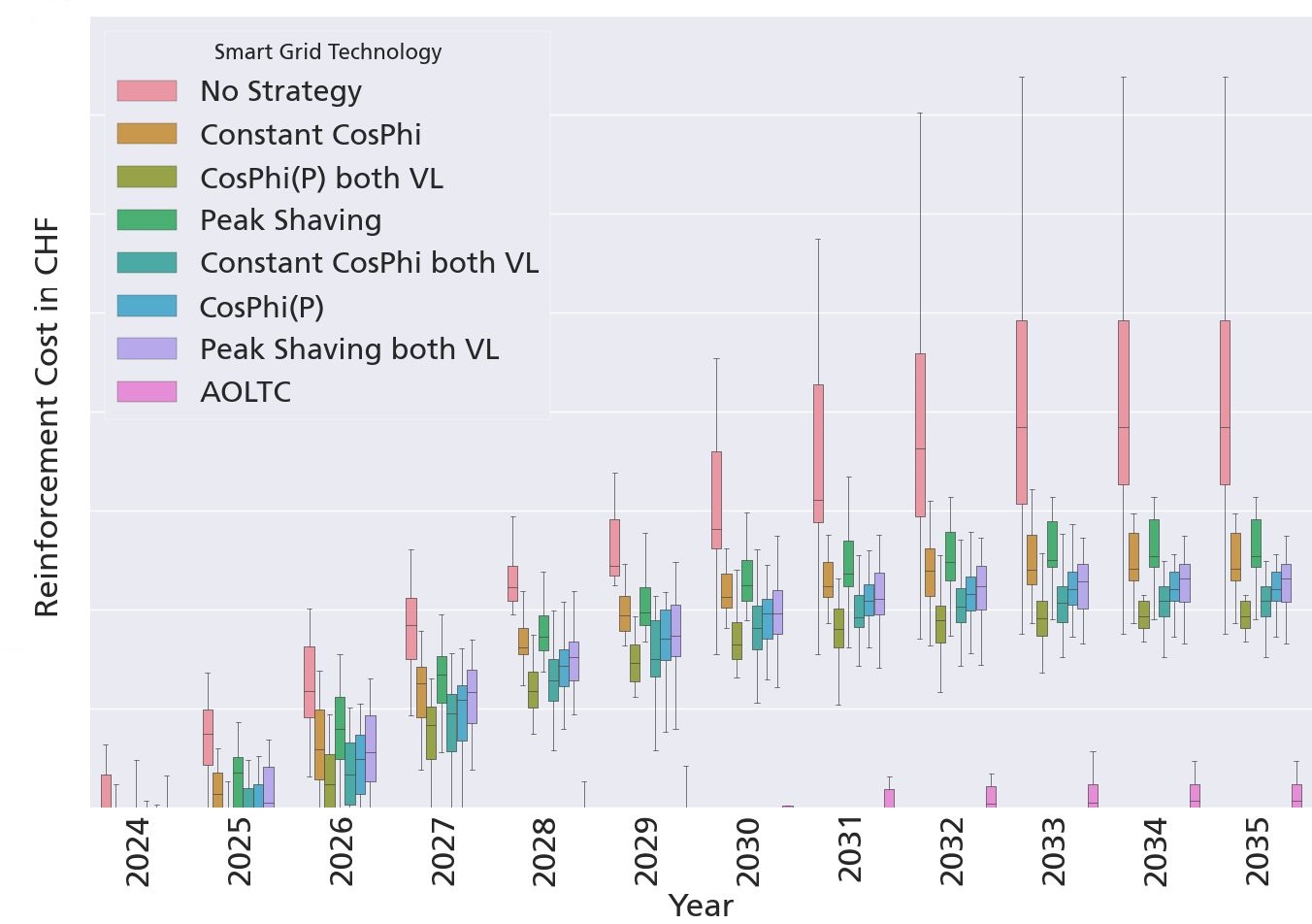}%
\label{romande_mv}}
\\
\subfloat[Comparison of avoided network reinforcement costs and OLTC investment costs for 84 LV networks of the German DSO Bayernwerk]{\includegraphics[width=\linewidth]{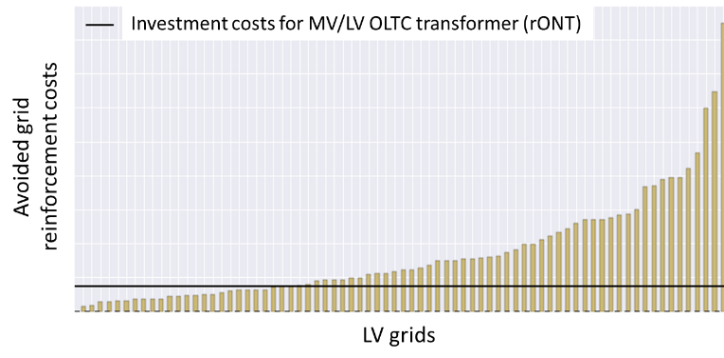}%
\label{fig_bag}}
\caption{Example results for the impact of Smart Grid technologies on reinforcement cost}
\label{fig_romande}
\end{figure}

The analysis of the PV hosting capacity of one MV network showed need for network reinforcement. It was therefore further investigated how control strategies can mitigate the expected reinforcement cost. The considered control strategies are:
\begin{itemize}
\item Constant CosPhi: reactive power provision by PV plants with a constant power factor
\item CosPhi(P): reactive power provision by PV plants with a power factor depending on active power provision
\item Peak Shaving: active power curtailment of PV plants
\item AOLTC: advanced on-load tap changer control of the HV/MV transformer, where the voltage set point of the transformer control is adapted to the active power flow over the transformer
\end{itemize}

Exemplary results are shown in Figure \ref{romande_mv}. The distribution of cost values is the result of the simulation of different distributions of PV systems in the networks.

It is clearly visible that all studied autonomous voltage control strategies are capable of reducing the expected reinforcement costs considerably. Due to the specific structure of the MV network mainly voltage rise is the limiting factor. Therefore, increasing the allowed voltage band by using an AOLTC has a significant effect on the expected reinforcement costs. It can reduce the PV integration cost by 90\% in 2035 compared with no strategy on average.

\subsection{Technical and Economical Assessment of MV/LV OLTC transformer}

In LV networks the application of an OLTC for the MV/LV transformer is a promising technical solution to increase the PV hosting capacity. In a network integration study in cooperation with the German DSO Bayernwerk the technical and economic potential of the OLTC was analysed for 84 real LV networks. The required network reinforcement costs for PV integration are determined with and without OLTC. Figure \ref{fig_bag} shows a comparison of the avoided network reinforcement costs by OLTC application and the OLTC investment costs for the LV networks. The OLTC is a cost-effective measure for PV integration for more than two thirds of the LV networks. It should be noted, that a high PV penetration scenario was investigated (PV rooftop potential) and that the operational costs are not considered in this example.

\subsection{Network Topology Optimisation}
A goal in strategic network planning can be to renew specific network elements, for example elements of a particular type (e.g. overhead lines or cables with a specific insulation) or elements that will have exceeded their life expectancy at the planning horizon. Consequently, sometimes large parts of a network area are to be renewed. In such a case it can be more cost-efficient to find a network structure with new line trails than simply maintaining the existing structure by renewing old line trails. 

\begin{figure}[!htt]
    \centering
    \subfloat[Initial network with lines \newline to remove (dotted)]{\includegraphics[width=0.48\linewidth]{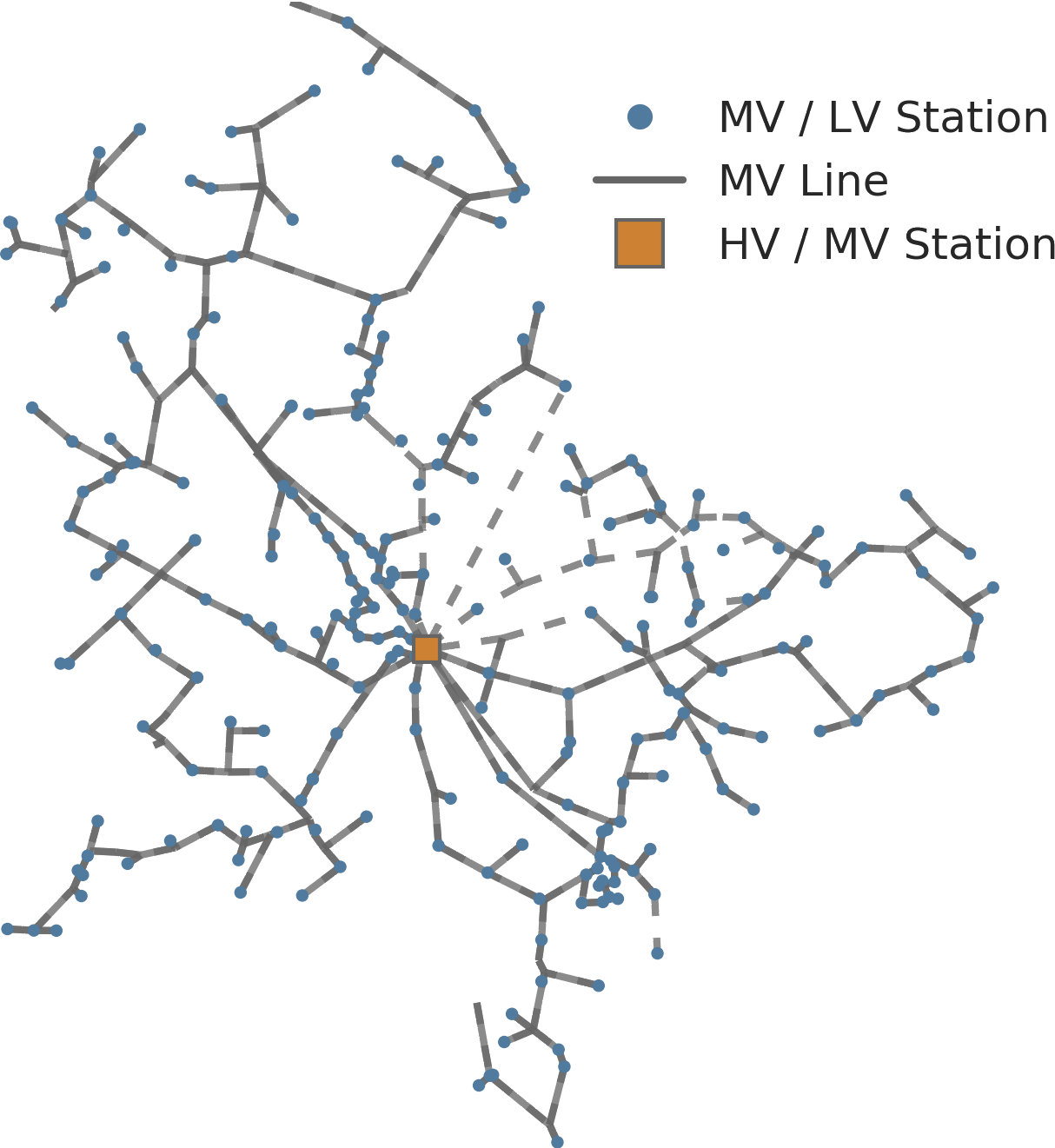} \hspace{0.15cm} 	\label{top_removed}}
    \subfloat[Line trails considered for new network structure]{\includegraphics[width=0.48\linewidth]{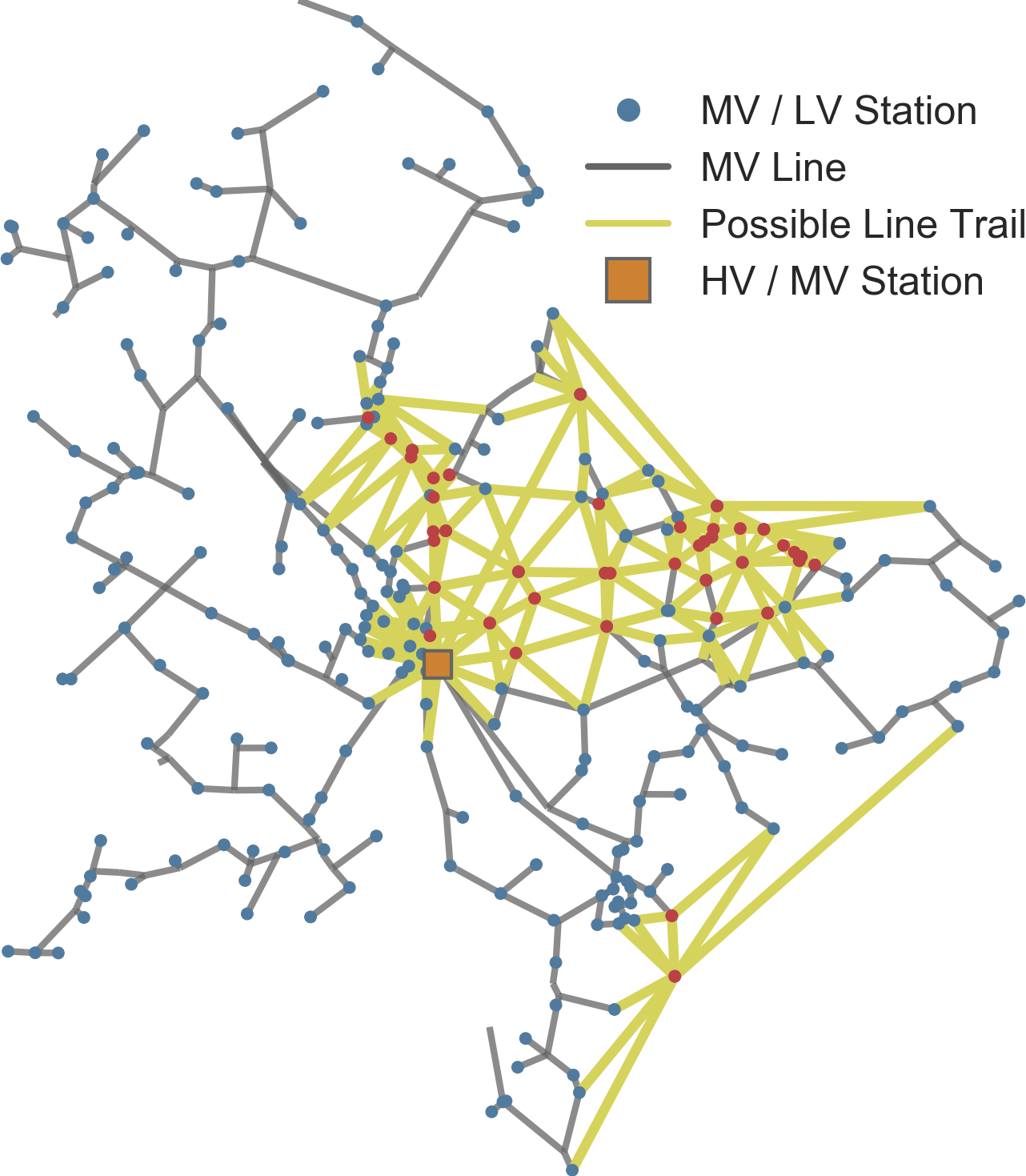}\hspace{0.15cm} \label{top_considered}} \\
    \subfloat[Optimized network \newline structure with new line trails]{\includegraphics[width=0.48\linewidth]{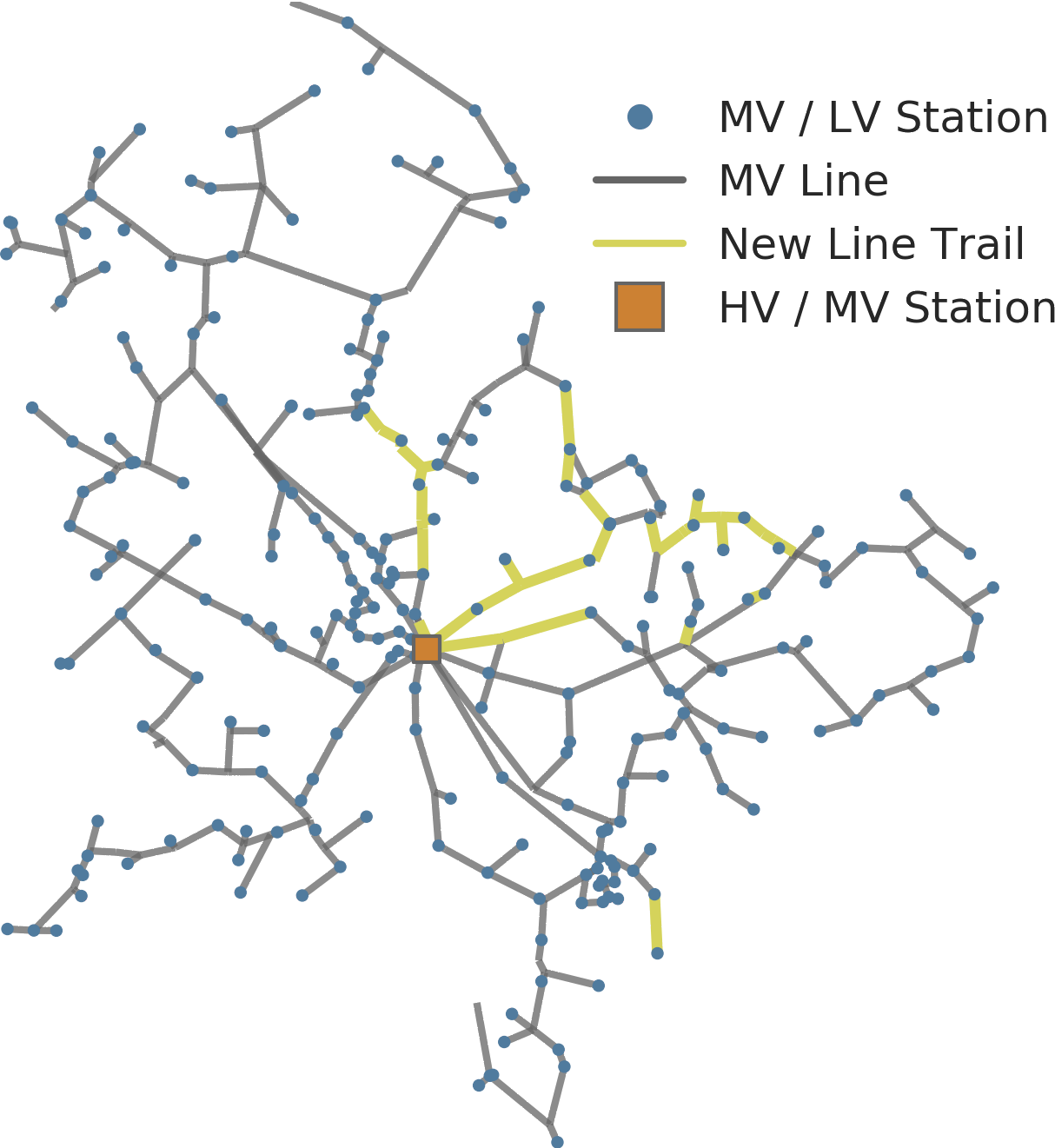} \hspace{0.15cm} \label{top_trails}}
    \subfloat[Feeder sectioning in optimized network structure]{\includegraphics[width=0.48\linewidth]{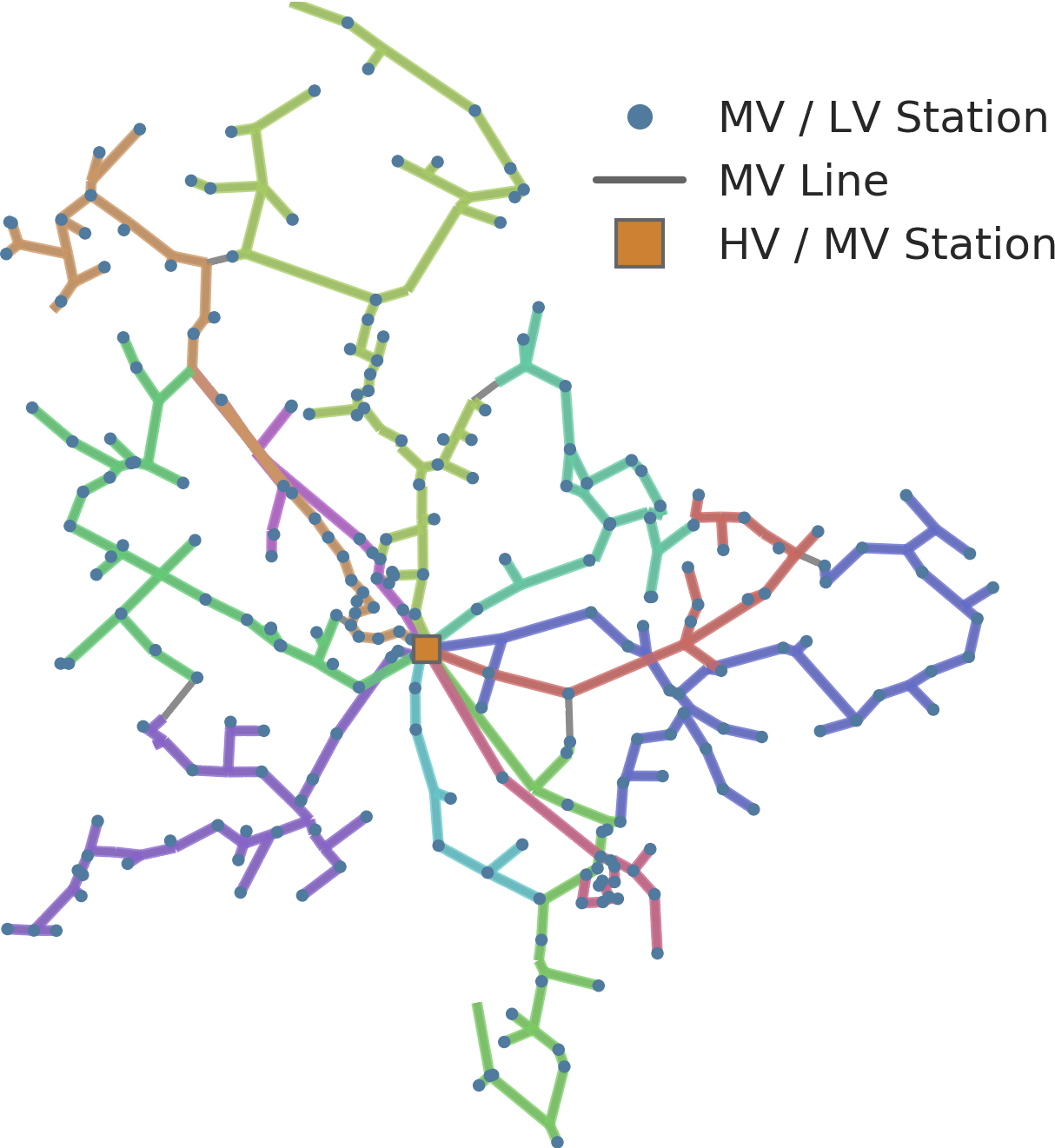} \hspace{0.15cm} \label{top_feeders}}
	\caption{Case study for topological network optimization}
	\label{topology}
\end{figure}

Figure \ref{topology} shows an example of the automated topology optimisation in a 10 kV network of the DSO Westnetz GmbH that has been developed in the scope of the project ANaPlan. The lines that are to be renewed in the target network are shown as dashed lines in Figure \ref{top_removed}. Figure \ref{top_considered} shows the possible new line trails that are considered in the optimization. Each line trail is modelled as one network reinforcement measure as explained in Section \ref{nep}. The line length of a new trail is assumed to be equal to the airline distance multiplied with a factor of 1.5 to account for obstacles. Additionally, switch measures are considered in the optimisation to allow for a reconfiguration of the feeder partitioning. In this example, 232 line trail measures and 605 switching measures have been considered. With this kind of complexity, it is important to use problem specific knowledge (like switches on stubs can never be opened, two switches on the same feeder sections can never be opened at the same time etc.) to restrict the problem. The automated network optimisation is then used to find a solution that requires a minimum amount of cabling while complying with all defined constraints. In this example, topological and operational constraints are used. The topological constraints ensure the radial structure of the network and possibilities of resupply in case of line faults. The operational constraints ensure the compliance with voltage band and maximum line loading in worst-case situations. The worst-case is modelled by one low load and high RES scenario and one high load and low RES scenario. By applying a prognosis for load development and RES installation at the planning horizon, the methodology can be used to find cost-efficient network structures for future power systems. The best solution that is found by the optimisation can be seen in Figure \ref{top_trails}. The feeder configuration in Figure \ref{top_feeders} shows that the radiality of the network is maintained. The optimized network structure leads to about \unit[25]{km} of new line trails, while a renewal of the old line trails would have resulted in about \unit[31]{km}. The optimisation was therefore able to find a more efficient structure that complies with topological and operational constraints considering load and RES development.

\section{Conclusion}
\label{conclusion}
This paper introduces an approach for automatic network planning that is a core component of our framework for large-scale network studies. It allows the calculation of network reconfiguration, reinforcement and extension with detailed network models. Since no data reduction or simplification is necessary, results of the automated network planning can be directly compared to solutions of experienced network planners to validate the results and improve the algorithm. This increases the transparency of RES network integration studies and permits direct conclusions regarding the network planning principles of the DSO. A further benefit of the automated approach is the possible application of probabilistic load and generation scenarios in the network integration study, which further increases the robustness and informative values. The network integration studies can be performed for a large number of real distribution networks, which avoids inaccuracies caused by the simplification of characteristic network models and their projection on a system-wide perspective. Therefore, the approach covers the complete diversity of the investigated distribution networks and allows conclusions for the network planning process of the individual network sections. The case studies presented in this paper highlight the need for detailed investigations, since results vary greatly in different case studies with different DSOs.

There are two main challenges for automated network studies. The first is the availability of data: quality and the informative value of the RES network integration studies are strongly dependent on the provided data base. Since a detailed knowledge of network conditions will become increasingly important in active distribution networks, most DSOs are actively working on improving data maintenance and standardisation to facilitate automated data analysis. An improvement in data quality and availability is thus to be expected.

Another challenge is the complexity of the network planning optimisation problem, which rises exponentially with the number of considered measures. Metaheuristic optimisation has been shown to find good solutions even for heterogeneous problems with several hundreds of measures.
The presented algorithms have been validated and improved with the feedback of network planning experts of several DSOs. In future, the algorithm could not only be used for studies but also applied as a direct supporting tool in the network planning process.

\section{Acknowledgment}
This research was partly supported by the German Federal Ministry for Economic Affairs and Energy and the Projektträger Jülich GmbH (PTJ) within the framework of the projects \textit{SmartGridModels} (FKZ 0325616) and \textit{ANaPlan} (FKZ 0325923B); and by the Federal Ministry of Education and Research (BMBF) within the framework of the project \textit{ENSURE} (FKZ 03SFK1N0).

\bibliographystyle{iet}
\bibliography{bibtex}{}

\end{document}